\documentstyle[10pt,aaspp4]{article}
\tightenlines

\def\etal{{\em et~al}.}
\def\lya{Ly$\alpha$}
\def\lyb{Ly$\beta$}
\begin{document}

\title{Possible High-Redshift, Low-Luminosity AGN Activity\\
in the Hubble Deep Field}
\author{R. Michael Jarvis and Gordon M. MacAlpine}
\affil{Department of Astronomy, University of Michigan, Ann
Arbor, MI 48109}

\begin{abstract}
In the Hubble Deep Field (HDF), twelve candidate sources of
high-redshift (z$>$3.5) AGN activity have been identified. 
The color selection criteria were established by passing spectra
of selected quasars and Seyfert galaxies (appropriately
redshifted and modified for ``Lyman forest'' absorption), as well
as stars, observed normal and starburst galaxies, and galaxy
models for various redshifts through the filters used for the HDF
observations.  The actual identification of AGN candidates also
involved convolving a Laplacian-of-Gaussian filter with the HDF
images, thereby removing relatively flat galactic backgrounds and
leaving only the point-like components in the centers.  Along
with positions and colors, estimated redshifts and absolute
magnitudes are reported, with the candidates falling toward the
faint end of the AGN luminosity function.  One candidate has
been previously observed spectroscopically, with a measured
redshift of 4.02.  The number of sources reported here is
consistent with a simple extrapolation of the observed quasar
luminosity function to magnitude 30 in B$_{\rm Johnson}$. 
Implications for ionization of the intergalactic medium and for
gravitational lensing are discussed.
\end{abstract}


\twocolumn
\section{Introduction}

\begin{figure}[t]
\figurenum{1}
\plotone{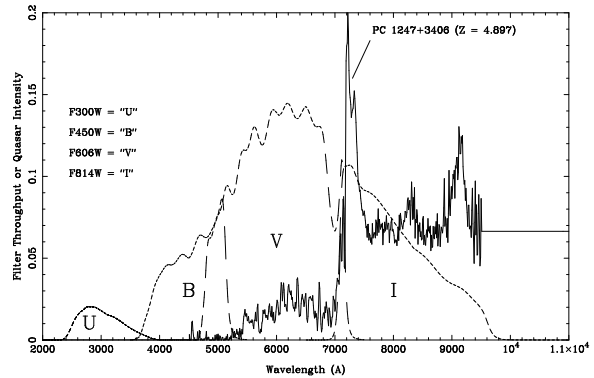}
\caption
{The net throughput for each of the four passbands used
in imaging the HDF.  We use the notation U, B, V, and I for the 
filters.  We also show a sample quasar spectrum to illustrate how 
the \lya\ emission and Lyman ``forest'' absorption yield distinctive 
colors as they pass through the different filters.
\label{fig1}
}
\end{figure}

The Hubble Deep Field (HDF) is a Director's Discretionary program
on the Hubble Space Telescope, which provides the deepest image
available for investigating galaxies at very high redshifts (see
Williams \etal\markcite{w2} 1996).  The field has an area of about 5 square
arcminutes, and it was selected for having low extinction and no
bright foreground sources.  Combined images were obtained with
the WFPC-2 camera through four filters: F300W, F450W, F606W, and
F814W.  For brevity, we call them U, B, V, and I, respectively. 
The net efficiency curves for the bandpasses are illustrated in
Figure~\ref{fig1}, taking into account both the filter throughput and the
quantum efficiency of the CCD detector.  Also shown is an example
of a high-redshift quasar spectrum superimposed on the filter
bandpasses.

The multicolor HDF images provide a unique opportunity for
investigating the faint end of the AGN luminosity function near
the epoch of galaxy formation, by identifying point sources
within galaxies and with AGN-like colors.  Finding pervasive,
comparatively low-level AGN sources could be quite important for
a variety of reasons.  For instance, Steidel and Sargent\markcite{s6} (1989)
postulated the existence of many low-luminosity AGN during the
epoch of galaxy formation, in order to explain ionization of
the intergalactic medium at high redshift
(see Gunn and Peterson\markcite{g1} 1965).  In this regard, 
Rees\markcite{r1} (1993) has
pointed out that, if intrinsically bright quasars are responsible
for intergalactic medium ionization, then they may influence or
modulate the development of large-scale environmental
inhomogeneities in the early Universe.  On the other hand, the
situation would be different for more smoothly distributed,
comparatively low-level AGN sources.  Also, Narayan\markcite{n1} (1989) noted
that a possible gravitational lensing explanation of reported
quasar-galaxy correlations could be facilitated by large numbers
of faint AGN at high redshift. 

Extrapolation of observed quasar surface densities (e.g.,
MacAlpine \etal\markcite{m2} 1988; Zamorani \etal\markcite{z1} 1991) to magnitude 30 in
B$_{\rm Johnson}$ for the area of the HDF suggests the likelihood
of finding approximately 3 to 30 objects.  For a redshift of
about 4, this apparent magnitude corresponds to an absolute
magnitude of roughly -16, near the low end of the class 1 Seyfert
galaxy luminosity function.  Although this simple process
involves a huge extrapolation in magnitude and thus should not be
taken too seriously, it provides a crude first estimate of what
one might expect. 

We note that Elson \etal\markcite{e1} (1996) examined unresolved objects in
the HDF with I~$<$~28 and found no convincing very-high-redshift
quasar candidates.  Because of our search technique, involving
convolution with a Laplacian-of-Gaussian filter, we were able to
identify additional unresolved sources within host galaxy
backgrounds.  None of the AGN candidate sources reported here 
are in the Elson \etal\ list, as discussed in Section 4.

A number of groups have derived photometric redshifts for
galaxies in the HDF.  One of these studies, by Lanzetta \etal\markcite{l1}
(1996), reported 54 galaxies with z~=~3~-~4 and 36 with z~$>$~4,
including 4 with z~$>$~6.  In our investigation, we did not find
what we consider to be convincing evidence for AGN at z~$>$~5. 
In addition, we note that, when we determined the colors for some
of the previously reported very-high-z candidate galaxies using
only their high S/N central regions, we obtained values
consistent with relatively low-redshift galaxies.  Only when we
expanded the radius of integration for each galaxy, leading to
very low S/N characteristics, did the galaxy colors move into the
extremely high-redshift region of the color-color plane.  We do
not consider those colors to be reliable.

The results presented in this paper represent a continuation of
work previously reported by Jarvis \etal\markcite{j2} (1996).  Since that
time, we have considerably refined our technique for picking out
AGN candidates.  As noted above and discussed below, we now use a
Laplacian-of-Gaussian filter to find star-like sources within
galaxies, whose colors are then investigated.  This technique
produces significantly more reliable identifications, compared
with our old method of finding AGN within galaxies by simply
looking for compact groups of pixels that have quasar-like
colors.

\section{Selection Criteria}

Because the HDF was observed through four filters, up to three
independent colors may be employed in the selection criteria.  
To determine what criteria are most useful for identifying AGN,
we investigated the colors for many different objects:
specifically stars, observed galaxies including starbursts,
modeled galaxies, quasars, and Seyfert galaxies.  Ultimately, we
decided not to use the U filter in the bulk of this study and to
concentrate on the B-V and V-I colors.  The reasons for this
decision are twofold.  First, we could not get reliable spectral
data in that passband for most of the objects of interest, and
thus could not accurately determine the predicted U-B colors
(for instance).  Second, the stars from which we calibrated our
point-source magnitudes (see below) were quite dim in U, so we
could not get a very good fit for this bandpass, and thus the
values for these magnitudes are not very reliable.  Because of
Lyman line and continuum absorption, we would expect weak U-band
fluxes for high-z AGN sources.  We retroactively test our 
candidates for this after the identification process.

For each type of object used in establishing the color selection
criteria, we considered measured spectra whenever possible and
theoretical or extrapolated spectra when necessary due to limited
spectral coverage of the observed data.  We integrated each
spectrum weighted by the throughput for each filter.  Then,
effective AB magnitudes were calculated using 
$m_{AB}~= \log(F)~+$ constant,
where $F$ is the constant flux that would give the same
integrated light as the actual spectrum, and the constant is
determined from the published zero points of the WFPC-2 filters.  A
summary of the results may be seen in the color-color plot of
Figure~\ref{fig2}, and more information about the spectra is presented
below.

\subsection{Stars}

For sample stars, we used actual spectra observed by Gunn and
Stryker\markcite{g2} (1983).  Their data include spectra of main sequence
stars from O5 to M8 plus a range of giants and supergiants.  The
most important for this study are the K and M dwarfs since they
begin to approach the region of color space where high-z AGN are
found.  Specifically, they have B-V colors approaching 2.

\subsection{Observed Galaxies}

For some of the galaxy colors, we used spectra of nearby galaxies
observed by Kennicutt\markcite{k1} (1992).  Because these data do not extend
far enough toward the blue for our purposes, we added appropriate
extrapolations toward shorter wavelengths, as derived from models
by Worthey\markcite{w3} (1994).  Artificially redshifting the spectra, we
found their resulting colors for redshifts up to 0.3.  Above
this, the colors are probably not accurate (see, for example,
Dorman \etal\markcite{d2} 1993), and modeled galaxies may be expected to
provide more appropriate spectra.

\begin{figure}[t]
\figurenum{2}
\plotone{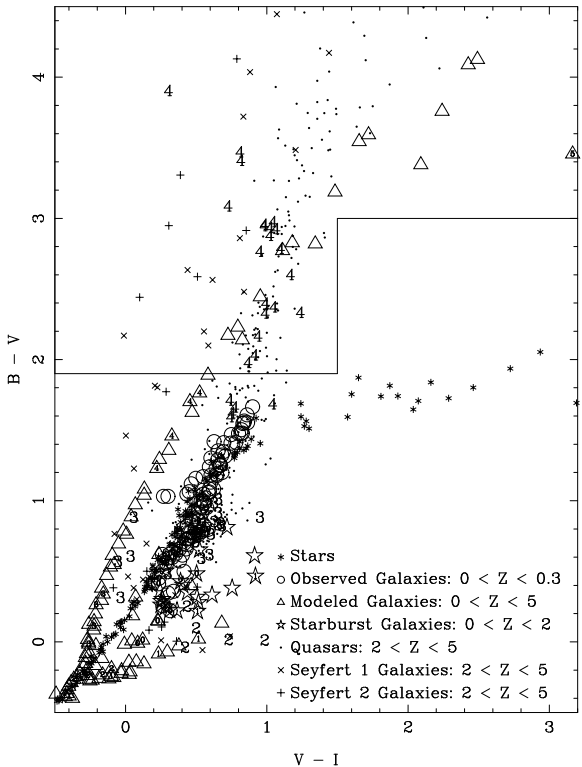}
\caption
{Expected colors for different kinds of objects at
various synthetic redshifts.  Redshift values are incremented
by 0.2 in all cases.  The solid line delimits the region above
which z~$>$~3.5 AGN candidates would lie. Note that this line 
excludes the highly-reddened starburst galaxy colors illustrated
in Figure~\ref{fig3}, which explains its shape.
\label{fig2}
}
\end{figure}

\subsection{Modeled Galaxies}

For very high redshifts, we considered 
modeled (stellar population synthesis) galaxy spectra.  We chose
to use the Bruzual and Charlot\markcite{b2} (1993) constant star formation
models for ages of 10$^{6}$, 10$^{7}$, 10$^{8}$, and 10$^{9}$ years.

The spectra and colors of high-redshift objects are influenced by
absorption due to intergalactic neutral hydrogen.  Whereas most
of the hydrogen is ionized, enough exists in the neutral state to
cause significant Lyman $\alpha$ absorption (the Lyman $\alpha$
forest), along with weaker Lyman $\beta$, and Lyman continuum
absorption.  Using the technique originally employed by Oke and 
Korycansky\markcite{o2} (1982), Zuo \etal\markcite{z3} (1993) 
characterized this effect with three parameters:
$D_{A}$, $D_{B}$ and $D_{C}$, corresponding to
Lyman $\alpha$, $\beta$ and continuum (hereafter \lya, \lyb\ and
Lyc) absorption, respectively.  Their best fit for $D_{A}$ involves a
formula that changes at some redshift which they call $z_{B}$. 
In their predicted functional form for $D_{A}$, there are 4
parameters: $z_{B}$, $\gamma_1$, $\gamma_2$, and $B$.  They did a
least squares fit for these parameters based on 32 QSO spectra
and obtained the following results:

     for $z < z_{B}, \ \ \ \ln (-\ln (1-D_{A})) = \ln B +
(\gamma_{1}+1) \ln (1+z)$; 

     for $z > z_{B}, \ \ \ \ln (-\ln (1-D_{A})) = \ln B +
(\gamma_{1}-\gamma_{2})
 \ln (1+z_{B}) + (\gamma_{2}+1) \ln (1+z$),

\begin{flushleft}where the best fit parameters are $\ln B =
-6.61$, $z_B = 3.11$, $\gamma_{1} = 2.82$, and $\gamma _2 =
5.07$.  
\end{flushleft}

One way to understand the physical meaning of $D_{A}$ is that the
optical depth due to the \lya\ forest is $-\ln (1-D_{A})$. 
Alternatively, $1-D_{A}$ is the average value of the observed
flux between \lyb\ and \lya, divided by the continuum flux level
just longward of \lya.  Whereas \lya\ absorption has by far the dominant
impact on the broadband colors, we have also included associated  
\lyb\ and Lyc absorption at relevant frequencies.

Taking these absorption effects into account, the modeled galaxy 
spectra for each age were
artificially shifted to redshifts between 0 and 5, with the
resulting colors shown in Figure~\ref{fig2}.  Numbers inside the 
triangles indicate integral values of z.

\begin{figure}[t]
\figurenum{3}
\plotone{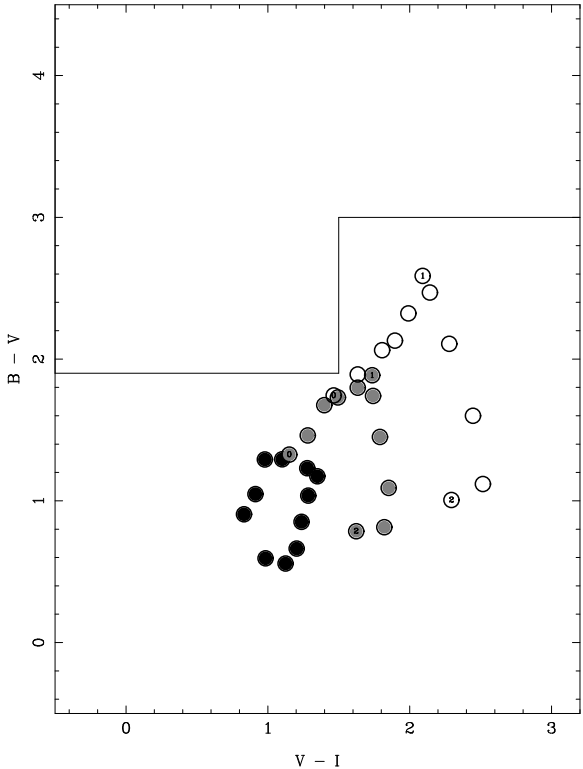}
\caption
{A starburst nucleus galaxy spectrum has been given 
additional reddening and red- shifted from 0 to 2, increment- ed by 
0.2, with resulting colors shown. The black cir- cles represent 
the original spectrum with an additional c~=~0.5 reddening.  
Grey and white circles represent c~=~1.0 and c~=~1.5 additional
reddening, respectively.  Numbers are given in the circles for 
integral redshift values.
\label{fig3}
}
\end{figure}

\subsection{Starburst Galaxies}

We considered actual spectra of the starburst nucleus galaxies
Mkn~357 and NGC~6090 (from the Astrophysics Data Facility IUE Archives),
with necessary extrapolations toward longer wavelengths
taken from the broad-band spectral energy distributions of
Schmitt \etal\markcite{s1} (1997).  Colors for each 
of these galaxies were plotted for incremental redshifts between 
0 and 2 in Figure~\ref{fig2}.  We could not meaningfully consider redshifts
above 2 due to lack of knowledge about the relevant spectral properties 
for wavelengths shortward of 1200~\AA.

We also investigated what colors could result if low to 
moderately-high redshift starburst galaxies have substantial 
internal dust obscuration.  Figure~\ref{fig3} shows three additional
examples for possible starburst galaxy spectral colors with
incremental redshifts from 0 to 2.  The black circles correspond
to the NGC~6090 spectrum (with additional data from Schmitt \etal), 
reddened using Osterbrock's\markcite{o1} (1989) average reddening law with c~=~0.5.
The grey and open circles correspond to the same galaxy
with c~=~1.0 and 1.5, respectively.  These are fairly extreme
amounts of reddening, but not impossible for a starburst galaxy. 
The numbers in some circles are integral values of z.  

\subsection{Quasars}

We used the 34 high-redshift (3.1~$<$~z~$<$~4.9) quasar spectra
observed by Schneider \etal\markcite{s2,s3} (1991a,b).  A power law
extrapolation of $F_{\nu}~\propto~\nu^{-2}$ (Zheng \etal\markcite{z2} 1997)
was inserted at the blue end where necessary.  Since the observed
spectra have effects of intergalactic hydrogen absorption for
their actual redshifts, we removed what was already there using
the above formulae and then put it back for the appropriate
synthetic redshift in each case.  On Figure~\ref{fig2}, we plotted the
quasars' colors as artificially redshifted to values of z within
$|\Delta z| \leq 1$ of the observed redshifts, 
with colors designated by isolated numbers at integral 
redshift values.  

\subsection{Seyfert Galaxies}

We considered both class~1 and 2 Seyfert galaxies.  Seyfert~1
spectra include Mkn~509, NGC~3783 and NGC~5548, as reported
by Koratkar \etal\markcite{k2} (1997).  For Seyfert~2 galaxies, we used
spectra of Mkn~573 (see MacAlpine\markcite{m1} 1988) and Mkn~3
(from the Astrophysics Data Facility IUE Archive).  In all cases,
geocoronal \lya\ emission was removed if necessary, an 
$F_{\nu}~\propto~\nu^{-2}$ continuum extrapolation was added for 
$\lambda~<~1200~$\AA, 
absorption effects of intergalactic hydrogen were
included as discussed above, and colors were plotted for
incremental redshifts between 0 and 5 in Figure~\ref{fig2}.

The Seyfert~1 and 2 locations in the color-color plane do not
differ substantially, and both are only moderately shifted to
higher B-V and lower V-I compared with the quasars.  The latter
may be understood as a luminosity effect, whereby
lower-luminosity AGN have higher line equivalent widths for a
given redshift (Baldwin\markcite{b1} 1997) and also (possibly) partly as an
evolution effect involving stronger line cores for lower-redshift
objects (see Francis and Koratkar\markcite{f1} 1995).

\bigskip

{\em Based on Figures~2 and 3, color selection criteria  
for identifying AGN-like spectra at redshifts from about 
3.5 to 5 are: B-V~$>$~1.9 and V-I~$<$~1.5,
or B-V~$>$~3 and V-I~$>$~1.5, as delineated by lines on the
figures.  Part of this region of color space may also be expected
to contain z~$>$~4 galaxies according to the Bruzual 
and Charlot models.}

\begin{figure}[b]
\figurenum{4}
\plotone{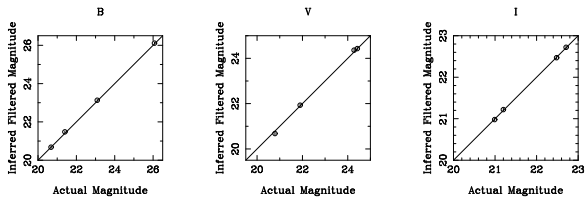}
\caption
{The inferred B, V and I magnitudes of the calibrating
stars versus their actual mag- nitudes given by 
Williams \etal\ (1996).  The lines shown have a slope 
of 1 and indicate where the stars should fall if the 
calibration were perfect.
\label{fig4}
}
\end{figure}

\section{HDF Data Analysis}
Since AGN are characterized as being galaxies with star-like
sources in the centers, we convolved a second derivative filter
with the images to enhance the contrast in the data before
looking at colors.  Specifically, we used a
``Laplacian-of-Gaussian operator'' (e.g., Jain \etal\markcite{j1} 1995), also
known as a ``sombrero'' filter.  It is equivalent to smoothing
the image with a Gaussian filter, and then taking the Laplacian
($\nabla ^{2}$, the two-dimensional second derivative) of the
result.  The scale size for the smoothing part of the function
was taken to be the radius of the point spread function (PSF) for
the HDF, which is optimal for identifying features of this size. 
The functional form of the operator is 
$(2-r^{2})\exp(-r^{2}/2)$, 
where $r$ is a radial distance measured in units
of $r_{0}$, the PSF radius.  The value of $r_{0}$ was derived
from 4 stars in the field and was found to be 1.33 pixels (0.053
arcsec).

The result of the operation described above is that the contrast
of star-size features is greatly magnified.  This includes
features which are small in only one direction, like spiral arms. 
The smooth parts of the galaxies are reduced to noise, since the
second derivative is essentially zero.  Stars are converted into
islands, tall peaks surrounded by negative valued
troughs\footnote{An ``ideal'' star, for which the flux
distribution is $F~= (A/2\pi)\exp(-r^{2}/2)$ (where $r$ is again
measured in units of $r_{0}$, the PSF radius, and A is a
constant), takes the following form after being processed:
$F^{\prime}~= (A/8)(4-r^{2})\exp(-r^{2}/4)$.}.  Thus, AGN may
be expected to appear
morphologically as stars at the centers of where galaxies appear
in the unprocessed image.

Because of the nature of this operation, it does not make sense
to integrate the processed image to find magnitudes.  We do know,
however, that the height of a resultant ``island'' is
proportional to the intensity of the star-like component, so
actual stars would be good calibrators.  Therefore, we inferred
the magnitudes of the processed objects by comparing them to the
processed images of 4 stars in the field, whose magnitudes are
known\footnote{For these stars, we used the ``total'' magnitudes
reported by Williams \etal\markcite{w2} (1996).}.  A plot of the calibration
is given in Figure~\ref{fig4}, showing the inferred magnitudes of the
stars versus their actual magnitudes.  We estimate that the error
induced by this process is less than 0.1 in each color.

\begin{figure}[t]
\figurenum{5}
\plotone{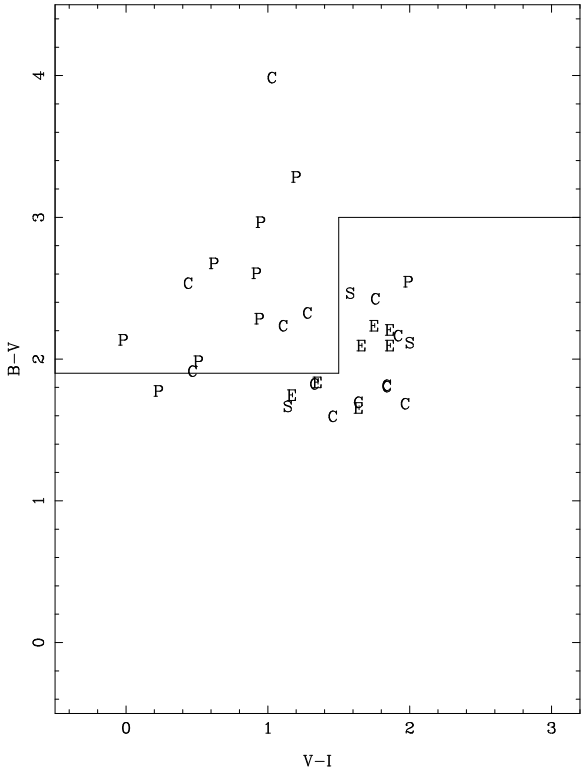}
\caption
{All HDF sources with processed B-V colors significantly
(meaning more than one sigma) greater than 1.5 are shown.  For 
reasons discussed in the text, objects above the solid line may 
be considered z~$>$~3.5 AGN candidates.  The letter P denotes a 
point source, less than 0.5 arcsec in diameter on the unprocessed 
image.  Objects plotted as a C for compact are between 0.5 and 
1.0 arcsec in diameter. E and S indicate larger ellipticals and 
spirals, respectively.  Some of the plotted B-V colors are actually 
lower limits, as indicated in Table~1.
\label{fig5}
}
\end{figure}

Figure~\ref{fig5} is a color-color plot for those objects with B-V 
above or near 1.9.  The high-z AGN criteria require a B-V color greater
than 1.9, but to avoid objects with a poorly defined color we
made the additional requirement that only those processed objects
with one-sigma B-V error bars that do not extend below 1.5
be included.  Of these 32 sources, 12 fall within the previously
defined, high-z AGN region of color space, as delineated by lines
in the figure.  The majority of other designated sources lie
where one could expect moderate-redshift, highly-reddened 
starburst galaxies, as illustrated in Figure 3.

Considering the colors of their ``point-like'' components,
we regard the 12 objects in the high-z AGN region
of color space as ``AGN candidates,'' even though Figure 2 indicates
that one might also expect to find some z~$>$~4 galaxies there.
We note that the majority of these sources (all those with B-V~$>$~2.5)
have significantly larger B-V measured for the processed point source
than for the host galaxy (from Williams \etal\markcite{w2} 1996), as would be
expected for an AGN source within a galaxy according to Figure~\ref{fig2}.
(B-V colors for galaxy models at z~=~4 lie below B-V colors
for AGN at z~=~4.)
Finally, we justify our reference to all of the 12 aforementioned objects
as AGN candidates because a major result of this work is to set
an {\em upper limit} to the number of such sources. 

The letters in Figure~\ref{fig5} indicate morphologies, which are somewhat
subjective and are based on the visual appearance of the
unprocessed image.  The letter P indicates that the unprocessed
object is not much bigger that a point source \-- specifically,
less than 0.5 arcsec in diameter.  The letter C is used for
compact galaxies which are clearly larger than point sources but
still less than one arcsec in diameter.  The letter E designates
an elliptical galaxy larger than one arcsec across.  Similarly, S
indicates a spiral galaxy with a diameter larger than one arcsec.

Figure~6 shows both the processed and unprocessed images of all
objects which fall above the line in the high-z AGN region of the
diagram, plus selected others which are intended to give the
reader a better idea of what the Laplacian-of-Gaussian filter
does.  Numbers next to the images correspond to ID numbers given
by Williams \etal\  The effect on spiral arms is especially
striking, as various features may be seen to stand out more in the
processed images.  Also, some ellipticals have steep enough cores
that the processed shape appears similar to that of a star.

Table~1 lists all 32 sources that are shown in Figure~\ref{fig5}.  It
contains each object's ID number, its HDF coordinate position
(frame, x, y), its sky position (RA, Dec), the colors and
magnitude of the point source (B-V, V-I, and I), the colors and
magnitude of the unprocessed galaxy (from Williams \etal\markcite{w2} 1996),
the measured redshift of the object if available (from Cohen
\etal\markcite{c1} 1996; Steidel \etal\markcite{s4} 1996; 
or Dickinson\markcite{d1} 1998), and the
observed morphology using the same notation as above.  For the
colors and magnitudes of the point sources, estimated one-sigma
errors are listed in parentheses.  For example, 1.28(12) means
1.28~$\pm$~0.12.  Errors are due to a combination of image noise
and calibration noise.  As mentioned above, we could not get
accurate measurements of U-B colors for the processed objects
because the calibrating stars were generally very faint in the U
band.  Therefore this color is listed, when available, for the host galaxies, 
but not for the point sources.

For each AGN candidate, Table~2 gives our rough estimate of the
redshift based on location in Figure~\ref{fig5}
and the implied absolute B magnitude of the host galaxy,
assuming $q_{0} = 0.5$ and $H_{0} = 75$ km s$^{-1}$
Mpc$^{-1}$.  Derived M$_{B}$ values for the point sources 
within the galaxies would be typically 1-2 magnitudes fainter.  

\section{Discussion}
As illustrated in Figure~\ref{fig5}, there are 12 sources which have the
morphology and colors to be considered z~$>$~3.5 AGN candidates.  
As a further consistency check, we looked at the U-B colors of the unprocessed
objects.  If these host galaxies really are at z~$>$~3.5, they
should also be Lyman break objects.  That is, they should not
have a measurable U-band magnitude.  All 12 of our candidates
were not visible in the U band according to the colors listed by
Williams \etal, so they are all considered to be viable z~$>$~3.5 AGN
candidates.  Precise redshift values are not possible for most of
them, but we list rough estimates in Table~2.

\begin{table}[t]
\begin{tabular*}{229pt}{l@{\hspace{44pt}}c@{\hspace{44pt}}@{\extracolsep{\fill}}c}
\multicolumn{3}{c}{Table 2.  Inferred Data For AGN Candidates} \\
\hline \hline
& Estimated &  \\
\hspace{10pt}ID & Redshift & M$_{\rm B}$$^a$ \\
\hline

2-282.0 &\phm{$>$}3.9 &\phm{$>$}-18.2 \\
2-436.0 &\phm{$>$}3.8 &$>$-17.6 \\
3-512.0$^b$ &\phm{$>$}3.7 &\phm{$>$}-19.2 \\
 &\phm{$>$}\em{4.0} &\phm{$>$}\em{-19.7} \\
3-675.0 &\phm{$>$}4.0 &\phm{$>$}-19.5 \\
3-775.0 &$>$4.0 &\phm{$>$}-18.0 \\
3-783.0 &$>$4.2 &$>$-18.4 \\
3-853.22 &\phm{$>$}3.7 &\phm{$>$}-17.5 \\
4-277.0 &\phm{$>$}4.4 &$>$-18.7 \\
4-280.0 &\phm{$>$}3.7 &$>$-17.5 \\
4-439.1 &$>$4.6 &$>$-19.4 \\
4-551.0 &$>$4.1 &$>$-17.8 \\
4-639.2 &\phm{$>$}3.8 &\phm{$>$}-18.5 \\
\hline
\end{tabular*}

$^a${\small Computed absolute magnitudes for 
$H_{0}$ = 75 km s$^{-1}$ Mpc$^{-1}$ and $q_{0}$ = 0.5.}     

$^b${\small The italicized values for this object are based
on the observed redshift.}
\end{table}

To our knowledge,
only one of our candidates has been observed spectroscopically.
As reported by Dickinson\markcite{d1} (1998), \#3-512.0 has a strong Ly$\alpha$ 
emission line for a measured redshift of 4.02.
This is reasonably close to our estimated redshift of 3.7 for this
object and lends credibility to our technique.   

In the Elson \etal\markcite{e1} (1996) list of unresolved HDF sources, none of
the objects correspond to AGN candidates reported here, as
noted in the Introduction.  However, their Table~1 does contain
four of our sources which have colors placing them just
outside the high-z AGN candidate region of Figure~5.  These are: \#3-419.0,
\#3-491.0, \#4-241.3, and \#4-713.0, all of which we designated as
``compact.''  (We note that the x and y pixel positions given by 
Elson \etal\ are offset from ours as well as from the standard system
of Williams \etal\ by +75 in each case.)
 
Absolute B magnitude estimates ($M_{B}$) in Table~2 for the host 
galaxies of the AGN
candidates were derived for the estimated redshifts, assuming
$H_{0}$~= 75~km~s$^{-1}$~Mpc$^{-1}$ and $q_{0}~=~0.5$.  We used
the expression 
$M_{B}~= B~+ 5\log(q_{0}^{2}H_{0})~- 5\log[q_{0}z~-
(1-q_{0})(\sqrt{1+2q_{0}z}-1)]~- k~- 52.39$
(derived from a luminosity distance given by Terrell\markcite{t1} 1977) where
k is the k-correction term.  As noted by Peterson\markcite{p1} (1997) (see
also Hewett\markcite{h1} 1992), k~$\approx$~0~mag for z~=~3.5
quasar spectra, and k~$\approx$~0.5~mag for z~=~4.  For our
calculations we used a linear fit to these values.  
The absolute magnitudes for AGN candidates 
in Table~2 represent the low-luminosity part of the expected range
for class~1 Seyfert galaxies.  Seyfert~1 absolute magnitudes typically
range from -22 to -18 with the faintest occurring at about -16
(see Veron-Cetty and Veron\markcite{v1} 1993).  The luminosity function for
Seyfert~2 galaxies has a maximum about a magnitude fainter than
for Seyfert~1 galaxies (Osterbrock\markcite{o1} 1989).

Our search found sources to a B magnitude of
roughly 30.  However, the list should not be considered
complete to this level.  The search technique is complex and
requires that the AGN be significantly brighter than the
surrounding galaxy light, so galaxies with large star forming
regions very close to the AGN could have been missed.  However,
we do not believe it is likely that we missed {\em many} of the
high-z AGN in the field, since the star-forming regions would
have to be {\em very} luminous, and/or the AGN sources would have
to be extraordinarily dim; and AGN candidates we did find already 
would represent unusually faint Seyfert galaxies.  
The ``expected'' number of AGN in
the field was about 10 plus or minus half an order of magnitude. 
Thus, the fact that we found 12 probable AGN suggests that there
is not a significant overabundance of faint, high-redshift AGN, as
postulated by Steidel and Sargent\markcite{s6} (1989) to account for
intergalactic medium ionization.

This result also sets strong limitations on the lensing
explanation for observed quasar-galaxy correlations (e.g.,
Webster \etal\markcite{w1} 1988).  Narayan\markcite{n1} (1989) noted that one way for a
lensing explanation to be consistent with the observed data is
for the faint end of the quasar/AGN luminosity function to be
significantly increased.  However, for this solution to work,
there would have to be about 300 AGN with I~$<$~28 in the HDF,
which is clearly ruled out by our result.  We believe, as did
Narayan, that a more likely explanation could involve a better
understanding of the completeness magnitude for the survey
used by Webster \etal

Of course, it must be born in mind that this investigation does 
not place limits on the number of AGN with z~$<$~3.5.  As illustrated
in Figure~\ref{fig2}, there are too many potential sources of confusion
in color space for this type of analysis
to be effective at lower redshifts.

\section{Summary}
Using color selection criteria and a Laplacian-of-Gaussian
filtering technique to identify point sources within galaxies, we
have searched the Hubble Deep Field for spectral/color
evidence of high-redshift, low-level AGN activity.  We identified
12 possible AGN sources with estimated 3.5~$\lesssim$~z~$\lesssim$~5. 
In all cases, estimated luminosities lie within the
low-luminosity range for Seyfert galaxies.  The results are
consistent with what could be expected from simple extrapolation
of the observed quasar luminosity function to very faint
magnitudes.  This study supports the view that faint AGN activity
does not play the major role in ionization of the early
intergalactic medium and also the view that observed
quasar-galaxy correlations do not imply a very large number of
faint AGN to be gravitationally lensed.

\acknowledgments
We would like to thank Gary Bernstein and Peter Jensen for
suggesting and assisting with the Laplacian-of-Gaussian process
in our search.  We also thank Guy Worthey for help in obtaining
galaxy spectra, Donald Schneider and Maarten Schmidt for sending
us digitized quasar spectra, Anuradha Koratkar for Seyfert galaxy
spectra and Charles Steidel for spectra of Lyman break objects
which contributed to improvements in the color determinations. 
Finally, we thank the referee for many useful comments which
helped generally to improve the paper.

Note: Table 1 is included as a separate file, since it is best viewed
in landscape format.  Figure 6 is also included as a separate file, since 
it is rather large.  Here is the caption for it:

\begin{figure}[h]
\figurenum{6}
\caption
{The processed and original images (on the left and
right, respectively) for each candidate AGN along with selected
other objects from Figure 5.  The first 12 are our AGN candidates.  
Each square is 1.7 arcseconds on a side, and the orientation for 
each CCD image is such that the x, y origin is in the lower left.  
\label{fig6}
}
\end{figure}

\newpage

\end{document}


\begin{deluxetable}{lccccccccccccccc}
\tablewidth{650pt}
\tablecaption{Measurements For Objects In Figure 5}
\tablecolumns{16}
\tablehead{
\colhead{Host} &
\multicolumn{3}{c}{Image Coords\tablenotemark{b}} &
\colhead{RA\tablenotemark{c}} &
\colhead{Dec\tablenotemark{d}} &
\multicolumn{3}{c}{Point Source Colors\tablenotemark{e}} & \colhead{} &
\multicolumn{4}{c}{Host Gal Colors\tablenotemark{f}} &
\colhead{} & \colhead{} \\
\cline{2-4} \cline{7-9} \cline{11-14} 
\colhead{Gal ID\tablenotemark{a}} &
\colhead{Fr} & \colhead{X} & \colhead{Y} &
\colhead{(12:)} & \colhead{(+62:)} &
\colhead{B-V} & \colhead{V-I} & \colhead{V} & \colhead{} &
\colhead{U-B} & \colhead{B-V} & \colhead{V-I} & \colhead{V} &
\colhead{Z\tablenotemark{g}} &
\colhead{Type\tablenotemark{h}} }
\startdata
\sidehead{AGN Candidates:}
2-282.0 &2 &\phn690 &\phn621 &36:49.03 &13:31.5 &\phm{$>$}2.32(70) &\phm{-}1.28(12) &28.27(8)\phn
& &$>$-0.65 &\phm{$>$}1.67 &1.20 &27.18 &\nodata &C\nl
2-436.0 &2 &\phn980 &\phn954 &36:50:10 &13:47.4 &\phm{$>$}2.28(61) &\phm{-}0.94(11) &28.16(8)\phn
& &\nodata &$>$1.62 &1.10 &27.70 &\nodata &P\nl
3-512.0 &3 &1304 &1146 &36:56.02 &12:45.7 &\phm{$>$}1.91(18) &\phm{-}0.47(9)\phn&26.96(7)\phn
& &\phm{-}$>$0.17 &\phm{$>$}1.67 &0.36 &25.85 &4.0 &C\nl
3-675.0 &3 &1369 &1460 &36:57.07 &12:35.3 &\phm{$>$}2.53(98) &\phm{-}0.44(17) &28.44(11)
& &$>$-0.16 &\phm{$>$}1.60 &0.46 &26.08 &\nodata &C\nl
3-775.0 &3 &\phn712 &1683 &36:54.14 &12:16.7 &$>$2.6\phn(11) &\phm{-}0.92(16) &28.46(13)
& &$>$-0.72 &\phm{$>$}1.67 &0.99 &27.53 &\nodata &P\nl
3-783.0 &3 &1604 &1705 &36:58.85 &12:30.1 &$>$3.0\phn(11) &\phm{-}0.95(12) &28.14(10)
& &\nodata &$>$2.07 &0.84 &27.04 &\nodata &P\nl
3-853.22 &3 &1550 &1835 &36:58.87 &12:24.5 &\phm{$>$}1.98(47) &\phm{-}0.51(12) &28.11(8)\phn
& &$>$-0.62 &\phm{$>$}1.99 &0.48 &27.19 &\nodata &P\nl
4-277.0 &4 &1837 &\phn604 &36:48.85 &11:53.9 &\phm{$>$}3.3\phn(10) &\phm{-}1.20(8)\phn&27.71(7)\phn
& &\nodata &$>$2.02 &1.07 &27.17 &\nodata &P\nl
4-280.0 &4 &\phn820 &\phn607 &36:46.58 &12:31.2 &\phm{$>$}2.13(49) &-0.02(14) &28.05(9)\phn
& &\nodata &$>$2.08 &0.29 &27.12 &\nodata &P\nl
4-439.1 &4 &\phn361 &\phn952 &36:43.75 &12:42.7 &$>$4.0\phn(11) &\phm{-}1.03(9)\phn&27.02(7)\phn
& &\nodata &$>$2.79 &1.07 &26.03 &\nodata &C\nl
4-551.0 &4 &1774 &1205 &36:45.56 &11:46.9 &$>$2.7\phn(11) &\phm{-}0.62(13) &28.42(11)
& &\nodata &$>$1.76 &0.40 &27.88 &\nodata &P\nl
4-639.2 &4 &\phn266 &1351 &36:41.45 &12:40.0 &\phm{$>$}2.23(40) &\phm{-}1.11(12) &27.72(10)
& &$>$-0.36 &\phm{$>$}2.26 &0.96 &26.18 &\nodata &C\nl
\sidehead{Other:}
2-977.0 &2 &\phn689 &1895 &36:55.68 &13:51.9 &\phm{$>$}1.77(21) &\phm{-}0.23(13) &27.43(8)\phn
& &$>$-0.20 &\phm{$>$}1.75 &0.46 &26.60 &\nodata &P\nl
3.266.0 &3 &1339 &\phn643 &36:55.06 &13:04.7 &\phm{$>$}2.42(15) &\phm{-}1.76(9)\phn&26.25(7)\phn
& &\phm{-}$>$0.50 &\phm{$>$}1.79 &1.65 &25.22 &\nodata &C\nl
3.321.1 &3 &\phn373 &\phn712 &36:50.17 &12:46.8 &\phm{$>$}2.09(11) &\phm{-}1.66(9)\phn&25.03(7)\phn
& &\phm{-}$>$2.17 &\phm{$>$}1.90 &1.47 &22.91 &0.7 &E\nl
3-378.0 &3 &1506 &\phn887 &36:56.49 &12:58.4 &\phm{$>$}2.54(77) &\phm{-}1.99(9)\phn&28.15(8)\phn
& &\nodata &$>$1.78 &1.74 &27.37 &\nodata &P\nl
3.419.0 &3 &\phn980 &\phn981 &36:53.95 &12:46.6 &\phm{$>$}1.69(8)\phn&\phm{-}1.64(7)\phn&24.36(6)\phn
& &\phm{-}\phm{$>$}0.91 &\phm{$>$}1.33 &1.57 &24.28 &\nodata &C\nl
3-430.2 &3 &1481 &1009 &36:56.64 &12:53.5 &\phm{$>$}1.80(23) &\phm{-}1.84(10) &27.36(9)\phn
& &\phm{-}\phm{$>$}1.56 &\phm{$>$}0.87 &1.51 &25.65 &1.2 &C\nl
3-486.0 &3 &1224 &1090 &36:55.48 &12:46.5 &\phm{$>$}2.46(25) &\phm{-}1.58(11) &26.50(9)\phn
& &\phm{-}\phm{$>$}0.49 &\phm{$>$}0.66 &1.00 &22.98 &0.8 &S\nl
3-491.0 &3 &1928 &1112 &36:59.21 &12:56.9 &\phm{$>$}1.68(8)\phn&\phm{-}1.97(7)\phn&24.43(6)\phn
& &\phm{-}$>$1.45 &\phm{$>$}1.65 &1.94 &24.42 &\nodata &C\nl
3-586.0 &3 &\phn473 &1288 &36:52.00 &12:27.4 &\phm{$>$}2.16(29) &\phm{-}1.92(10) &27.41(8)\phn
& &$>$-0.20 &\phm{$>$}1.68 &1.74 &26.03 &\nodata &C\nl
3-790.1 &3 &1919 &1688 &37:00.46 &12:35.7 &\phm{$>$}1.83(12) &\phm{-}1.35(10) &24.91(8)\phn
& &\phm{-}\phm{$>$}2.35 &\phm{$>$}1.59 &1.14 &22.51 &0.6 &E\nl
3-815.1 &3 &1144 &1749 &36:56.55 &12:21.2 &\phm{$>$}2.09(13) &\phm{-}1.86(9)\phn&26.28(8)\phn
& &\phm{-}\phm{$>$}0.78 &\phm{$>$}1.83 &1.69 &24.28 &\nodata &E\tablebreak
4-241.3 &4 &\phn666 &\phn524 &36:46.68 &12:38.1 &\phm{$>$}1.59(8)\phn&\phm{-}1.46(7)\phn&23.81(6)\phn
& &\phm{-}$>$2.93 &\phm{$>$}1.56 &1.45 &23.69 &0.3 &C\nl
4-254.0 &4 &\phn404 &\phn535 &36:46.04 &12:47.6 &\phm{$>$}1.65(11) &\phm{-}1.64(9)\phn&25.64(7)\phn
& &\phm{-}\phm{$>$}1.22 &\phm{$>$}1.27 &1.52 &23.85 &\nodata &E\nl
4-471.0 &4 &1706 &1012 &36:46.42 &11:52.4 &\phm{$>$}1.74(10) &\phm{-}1.17(9)\phn&24.47(8)\phn
& &\phm{-}$>$2.15 &\phm{$>$}1.67 &1.04 &22.95 &0.5 &E\nl
4-493.0 &4 &\phn300 &1057 &36:43.07 &12:43.3 &\phm{$>$}2.20(12) &\phm{-}1.86(9)\phn&25.25(8)\phn
& &\phm{-}$>$1.41 &\phm{$>$}1.83 &1.65 &23.34 &0.8 &E\nl
4-713.0 &4 &\phn335 &1504 &36:40.80 &12:35.1 &\phm{$>$}1.82(9)\phn&\phm{-}1.33(7)\phn&25.41(6)\phn
& &\phm{-}$>$0.92 &\phm{$>$}1.76 &1.29 &25.39 &\nodata &C\nl
4-744.0 &4 &1716 &1532 &36:43.72 &11:44.0 &\phm{$>$}2.23(11) &\phm{-}1.75(9)\phn&24.73(8)\phn
& &\phm{-}\phm{$>$}1.70 &\phm{$>$}1.97 &1.58 &22.40 &0.8 &E\nl
4-775.0 &4 &1555 &1583 &36:43.09 &11:49.1 &\phm{$>$}2.11(34) &\phm{-}2.00(13) &27.32(11)
& &\phm{-}\phm{$>$}1.09 &\phm{$>$}0.98 &1.42 &23.87 &\nodata &S\nl
4-776.0 &4 &1922 &1604 &36:43.79 &11:35.3 &\phm{$>$}1.81(24) &\phm{-}1.84(9)\phn&27.56(8)\phn
& &\phm{-}\nodata &$>$2.33 &1.70 &26.64 &\nodata &C\nl
4-948.12 &4 &1592 &1955 &36:41.22 &11:41.9 &\phm{$>$}1.66(13) &\phm{-}1.14(10) &26.03(8)\phn
& &\phm{-}\phm{$>$}1.58 &\phm{$>$}1.54 &1.20 &23.79 &\nodata &S\nl
\enddata
\tablenotetext{a}{\small ID number for the host galaxy from the
Williams \etal\ catalog.}
\tablenotetext{b}{\small Frame and X, Y chip positions (measured 
     from the corner of each chip closest to the center of the array)
     for the center of the stellar object.}  
\tablenotetext{c}{\small Measured Right Ascension (2000.0) of the 
     stellar object.  All RA values start with 12 hours.}
\tablenotetext{d}{\small Measured Declination (2000.0) of the 
     stellar object.  All Dec values start with 62 degrees north.}
\tablenotetext{e}{\small Colors and V magnitude of the processed
stellar object found by comparison with known stars in the
field.}
\tablenotetext{f}{\small Colors and V magnitude of the
unprocessed galaxy taken from Williams \etal\ (1996).}
\tablenotetext{g}{\small Observed redshift value if available
from Cohen \etal\ 1996, Steidel \etal\ 1996, or Dickinson 1998.}
\tablenotetext{h}{\small Observed morphology of the unprocessed
object.  P indicates a nearly point source, C a galaxy less 
than~1\arcsec\ in diameter, E a large (ie. larger than~1\arcsec)
elliptical, and S a large spiral.}
\end{deluxetable}